\documentclass[preprintnumbers,prd,twocolumn,showpacs,amsmath,amssymb,
nofootinbib , superscriptaddress]{revtex4}

\usepackage{graphicx}
\usepackage{color}

\usepackage{bm}
\usepackage{amsfonts}

\newcommand{\beq}{\begin{equation}}
\newcommand{\eeq}{\end{equation}}
\newcommand{\bea}{\begin{eqnarray}}
\newcommand{\eea}{\end{eqnarray}}
\newcommand{\bseq}{\begin{subequations}}
\newcommand{\eseq}{\end{subequations}}

\newcommand{\Ref}[1]{(\ref{#1})}

\begin{document}

\title{Cosmology with nonminimal kinetic coupling and a power-law
potential}

\author{Maria A. Skugoreva}
\affiliation{Peoples Frendship University of Russia, Moscow 117198,
Russia}
\author{Sergey V. Sushkov}
\email{sergey_sushkov@mail.ru}
\affiliation{Institute of Physics, Kazan Federal University,\\
Kremlevskaya 18, Kazan 420008, Russia}
\author{Alexei V. Toporensky}
\affiliation{Sternberg Astronomical Institute, Moscow 119992,
Russia}

\begin{abstract}
We consider cosmological dynamics in the theory of gravity with the scalar field
possessing a nonminimal kinetic coupling to gravity,
$\kappa G_{\mu\nu}\phi^{\mu}\phi^{\nu}$,
and the power-law potential $V(\phi)=V_0\phi^N$. Using the dynamical system
method, we analyze all possible asymptotical regimes of the model
under investigation and show that for sloping potentials with $0<N<2$ there exists a quasi-de Sitter asymptotic $H={1}/{\sqrt{9\kappa}}$ corresponding to an early inflationary Universe.
In contrast to the standard inflationary scenario, the kinetic coupling inflation does not depend on a scalar field potential and is only determined by the coupling parameter $\kappa$.
We obtain that there exist two different late-time asymptotical regimes. The first one leads to the usual power-like cosmological evolution with $H=1/3t$, while the second one represents the late-time inflationary Universe with $H=1/\sqrt{3\kappa}$. This secondary inflationary phase depends only on $\kappa$ and is a specific feature of the model with nonminimal kinetic coupling.
Additionally, an asymptotical analysis shows that for the quadric potential with $N=2$ the asymptotical regimes remain qualitatively the same,
while the kinetic coupling inflation is impossible for steep potentials with $N>2$.
Using a numerical analysis, we also construct exact cosmological solutions and find initial conditions leading to the initial kinetic coupling inflation followed either by a ``graceful'' oscillatory exit or by the secondary inflation.
\end{abstract}


\maketitle

\section{Introduction}
In recent decades plenty of remarkable results has been achieved in the
observational cosmology including precise measurements of the Cosmic
Microwave Background (CMB) radiation \cite{CMB}, systematic observations of
nearby and distant Type Ia supernovae (SNe Ia) \cite{supernova}, study of baryon
acoustic oscillations \cite{BAO}, mapping the large-scale structure
of the Universe, microlensing observations, and many others (see, for example,
the review \cite{Observations}). These achievements have set new serious
challenges before theoretical physics and prompted many speculations mostly
based on phenomenological ideas which involve new dynamical sources of gravity
that act as dark energy, and/or various modifications to general relativity.
The spectrum of models, having been postulated and explored in recent years, is
extremely wide and includes, in particular, Quintessence \cite{quintessence},
$K$-essence \cite{Kessense}, Ghost Condensates \cite{Ghost},
Dvali-Gabadadze-Porrati gravity \cite{DGP}, Galileon gravity
\cite{Ggravity}, and $f(R)$ gravity \cite{fRgravity}
(see Refs. \cite{SahSta, PeeRat, Nob, CopSamTsu, CalKam, SilTro, Cli_etal,
AmeTsu} for detailed reviews of these and other models).

The most of phenomenological models represents various modifications of
scalar-tensor theories. Of particular interest are models allowing for
nonminimal couplings between derivatives of scalar fields and the curvature.
%In 1993 Amendola \cite{Amendola} has been considered the most general
%Lagrangian linear in the curvature scalar $R$, quadratic in $\phi$,
%and containing terms with four derivatives, such
%as $\kappa_1 R\phi_{,\mu}\phi^{,\mu}$,
%$\kappa_2 R_{\mu\nu}\phi^{,\mu}\phi^{,\nu}$ and others (see also
%\cite{Capozziello} for
%details). Note that generally the order of field equations in the theory with
%nonminimal derivative couplings is higher than two. However, for the special
%case $\kappa G_{\mu\nu}\phi^{,\mu}\phi^{,\nu}$
As was shown by Amendola \cite{Amendola}, a theory with derivative couplings
cannot be recast into the Einsteinian form by a conformal rescaling
$\tilde g_{\mu\nu}  = e^{2\omega} g_{\mu\nu}$. He also supposed that an
effective cosmological constant and then the inflationary phase can be recovered
without considering any effective potential if a nonminimal derivative coupling
is introduced. Amendola himself \cite{Amendola} investigated a cosmological
model with the Lagrangian containing the only derivative coupling term $\kappa_2
R_{\mu\nu}\phi^{,\mu}\phi^{,\nu}$ and presented some analytical inflationary
solutions. A general model containing $\kappa_1 R\phi_{,\mu}\phi^{,\mu}$ and
$\kappa_2 R_{\mu\nu}\phi^{,\mu}\phi^{,\nu}$ has been discussed by Capozziello
{\em et al} \cite{Capozziello}. They showed that the de Sitter spacetime is an
attractor solution in the model. Further investigations of cosmological and
astrophysical models with nonminimal derivative couplings have been continued in
\cite{kincoupl,Bruneton,Tsujikawa,Sus:2009,SarSus:2010, Sus:2012,
SusRom:2012}.

Note that generally the order of field equations in models with nonminimal
derivative couplings is higher than two. However, it reduces to second order in
the particular case when the kinetic term is only coupled to the Einstein
tensor, i.e. $\kappa G_{\mu\nu}\phi^{,\mu}\phi^{,\nu}$ (see, for example, Ref.
\cite{Sus:2009}).\footnote{It is worth noting that a general single scalar field Lagrangian giving rise to second-order field equations had been derived by Horndeski \cite{Horndeski} in 1974. The model with $\kappa G_{\mu\nu}\phi^{,\mu}\phi^{,\nu}$ represents a particular form of the Horndeski Lagrangian. Recent interest in second-order gravitational theories is also connected with the Dvali-Gabadadze-Porrati braneworld \cite{DGP} and and Galileon gravity \cite{Ggravity}.}

%Recently there has been renewed interest in second-order gravitational theories in
%connection with the Dvali-Gabadadze-Porrati braneworld \cite{DGP} and Galileon
%gravity [29, 30] (see also Refs. [31-40]).}
%Deffayet et al. [41] showed that the most general action in those theories is
%given by
In our recent works \cite{Sus:2009,SarSus:2010,Sus:2012} we have investigated
cosmological scenarios with with the nonminimal derivative coupling $\kappa
G_{\mu\nu}\phi^{,\mu}\phi^{,\nu}$, focusing on models with zero and constant
potentials. According to the parameter choices, we have obtained the variety of
behaviors including a Big Bang, an expanding universe with no beginning, a
cosmological turnaround, an eternally contracting universe, a Big Crunch, and a
cosmological bounce \cite{SarSus:2010}. However, the most interesting and
important feature we have found is that the non-minimal derivative coupling
provides an essentially new inflationary mechanism and naturally describe
transitions between various cosmological phases without any fine-tuning
potential. The inflation is driving by terms in the field equations responsible
for the non-minimal derivative coupling. At early times these terms are
dominating, and the cosmological evolution has the quasi-de Sitter character
$a(t)\propto e^{H_\kappa t}$ with $H_\kappa=1/\sqrt{9\kappa}$, where $\kappa$ is
a coupling parameter with dimension of ({\em length})$^{2}$. Note that the
estimations give $\kappa\simeq 10^{-74}$ sec$^2$ \cite{Sus:2012}. Later, in the
course of the cosmological evolution the domination of $\kappa$-terms is
canceled, the usual matter comes into play, and the Universe enters into the
matter-dominated epoch.

The scalar potential plays very important and, frequently, crucial role in
scalar-tensor theories of gravity. Could the potential drastically modify
cosmological scenarios  with the non-minimal derivative coupling found in models
with zero and/or constant potentials? In the present paper we study this problem for a
power-law potential $V(\phi)=V_0\phi^N$.

%--------------------------------------------------------------------
\section{Action and field equations}
%--------------------------------------------------------------------
Let us consider the theory of gravity with the action
\begin{equation}\label{action}
S=\int d^4x\sqrt{-g}\left\{ \frac{R}{8\pi} -\big[g^{\mu\nu} + \kappa G^{\mu\nu}
\big] \phi_{,\mu}\phi_{,\nu} -2V(\phi)\right\},
\end{equation}
where $V(\phi)$ is a scalar field potential, $g_{\mu\nu}$ is a metric, $R$ is the
scalar curvature, $G_{\mu\nu}$ is the Einstein tensor, and $\kappa$ is the
coupling parameter with dimension of ({\em length})$^2$.

In the spatially-flat Friedmann-Robertson-Walker cosmological model the action
\Ref{action} yields the following field equations \cite{Sus:2012}
\bseq\label{genfieldeq}
\bea
  \label{00cmpt}
  &&3H^2=4\pi\dot{\phi}^2\left(1-9\kappa H^2\right) +8\pi V(\phi),\\
  &&\displaystyle
  2\dot{H}+3H^2=-4\pi\dot{\phi}^2
  \left[1+\kappa\left(2\dot{H}+3H^2 +4H\ddot{\phi}\dot{\phi}^{-1}\right)\right]
\nonumber\\
  \label{11cmpt}
  && \ \ \ \ \ \ +8\pi V(\phi),\\
  \label{eqmocosm}
  &&(\ddot\phi+3H\dot\phi)-3\kappa(H^2\ddot\phi
  +2H\dot{H}\dot\phi+3H^3\dot\phi)=-V_\phi,
\eea
\eseq
where a dot denotes derivatives with respect to time, $H(t)=\dot a(t)/a(t)$ is
the Hubble parameter, $a(t)$ is the scale factor, $\phi(t)$ is a homogenous
scalar field, and $V_\phi=dV/d\phi$. It is worth noticing that Eq.
\Ref{eqmocosm} can be rewritten as
follows
\beq\label{eqmoint}
\big[a^3(1-3\kappa H^2)\dot\phi\big]\!\dot{\phantom{\phi}}=-a^3V_\phi.
\eeq
In the case $V(\phi)\equiv const$, when $V_\phi=0$, Eq. \Ref{eqmoint} can be
easily integrated:
\beq\label{intphi}
\dot\phi=\frac{C}{a^3(1-3\kappa H^2)},
\eeq
where $C$ is a constant of integration.

Note that equations \Ref{11cmpt} and \Ref{eqmocosm} are of second
order, while \Ref{00cmpt} is a first-order differential
constraint for $a(t)$ and $\phi(t)$. The constraint (\ref{00cmpt}) can be
rewritten as:
 \beq\label{constrphigen}
 \dot\phi^2=\frac{3H^2-8\pi
 V(\phi)}{4\pi(1-9\kappa H^2)},
\eeq
or equivalently as
\beq\label{constralphagen}
  H^2=\frac{4\pi\dot\phi^2+8\pi
  V(\phi)}{3(1+12\pi\kappa\dot\phi^2)}.
\eeq
Therefore, as long as the parameter $\kappa$ and the potential
$V(\phi)$ are given, the above relations provide restrictions for the
possible values of $H$ and $\dot\phi$, since they have to give rise to
non-negative $\dot\phi^2$ and $H^2$, respectively. Assuming the non-negativity of the potential, i.e. $V(\phi)\ge0$, we can conclude from Eqs. \Ref{constrphigen} and \Ref{constralphagen} that in the theory with the positive $\kappa$ possible values of $\dot\phi$ are unbounded, while $H$ takes restricted values. Vice versa, the negative $\kappa$ leads to bounded $\dot\phi$ and unbounded $H$. Hereafter we will suppose that $\kappa>0$.

%%%%%%%%%%%%%%%%%%%%%%%%%%%%%%%%%%%%%%%%%%%
\section{Dynamical system}
%%%%%%%%%%%%%%%%%%%%%%%%%%%%%%%%%%%%%%%%%%%
In order to find asymptotic regimes of the system \Ref{genfieldeq}  we introduce the following set of dimensionless variables
%\begin{subequations}
\bea
&&
x=\frac{8\pi\dot\phi^2}{6H^2(1+8\pi\kappa\dot\phi^2)},\quad
y=-\frac{8\pi\kappa\dot\phi^2}{2(1+8\pi\kappa\dot\phi^2)},
\nonumber\\
&&
z=\frac{8\pi V}{3H^2(1+8\pi\kappa\dot\phi^2)},\quad
v=\frac{\dot \phi}{\phi H}.
\label{def:xyzv}
\eea
%\end{subequations}
Generally, $x$ characterizes the kinetic energy, and $z$ characterizes the
potential energy of the scalar field, while $y$ is connected with non-minimal
kinetic coupling. Correspondingly, $z= 0$ if $V=0$, and $y=0$ if $\kappa=0$.

Using new variables, we can rewrite Eq. \Ref{00cmpt} as follows
\beq
x+y+z=1.
\eeq
The latter is a constraint for values of $x$, $y$ and $z$. Using this
constraint, we can exclude $y$ from subsequent relations.

Differentiating Eqs. \Ref{00cmpt}, \Ref{eqmocosm}, and $v=\frac{\dot \phi}{\phi H}$, we  obtain
\bea\label{xzv}
x'& = &2x\left[X(3-2x-2z)-Y\right],\\
z'& = &z\left[\beta v-2Y+4X(1-x-z)\right],\\
v'& = &v\left[X-Y-v\right].
\eea
where the prime means a derivative with respect to $\ln a$,\footnote{One has the following relation: $\frac{d}{dt}=H\frac{d}{d(\ln a)}$.} and the following notations are used:
\beq\label{def:betaXY}
\beta=\frac{\phi V_\phi}{V}, \quad
X=\frac{\ddot\phi}{\phi H},\quad Y=\frac{\dot H}{H^2}.
\eeq
The dimensionless parameter $\beta$ depends on the specific form of $V(\phi)$.
Hereafter we will discuss the power-law potential
\beq
V(\phi)=V_0\phi^N.
\eeq
In this case we have $\beta=N={\rm const}$.

To express $X$ and $Y$ via $x,y,v$ and $N$, we differentiate Eq. \Ref{00cmpt}, and divide the obtained relation by $\frac{3}{4\pi}H^3(1+8\pi\kappa\dot\phi^2)$. After some algebra we obtain
\beq
2X(3-2x-3z)-2Y(x+z)+N v z=0.
\eeq
Then, dividing Eq. \Ref{eqmocosm} by $\frac{3}{4\pi\dot\phi^2}H^3(1+8\pi\kappa\dot\phi^2)$, we can find
\beq
X(1-z)+2Y(1-x-z)+3(1-z)+\frac12N v z=0.
\eeq
Resolving this system with respect to $X$ and $Y$ yields
\bea
X & = & \frac{1}{\Delta}\left[\frac12 N v z(x+z-2)-3(1-z)(x+z)\right],\\
Y & = & \frac{1}{\Delta}\bigg[N v z(x+z-1)+3(1-z)(2x+3z-3)\bigg],
\label{Y}
\eea
where $\Delta=-9x(1-z)-11z+5z^2+4x^2+6$.
Substituting these relations into \Ref{xzv} we obtain finally the following
dynamical system:
\begin{widetext}
\begin{subequations}\label{dynsys}
\bea
x'& = &\frac{2x}{\Delta}\left[(\textstyle\frac12 N v z(x+z-2)-3(1-z)(x+z))(3-2x-2z)- N v z(x+z-1)-3(1-z)(2x+3z-3)\right],\\
z'& = &\frac{z}{\Delta}\left[N v \Delta-2N v z(x+z-1)-6(1-z)(2x+3z-3)+2(N v
z(x+z-2)-6(1-z)(x+z))(1-x-z)\right],\\
v'& = &\frac{v}{\Delta}\left[\frac12 N v z(x+z-2)-3(1-z)(x+z)-N v z(x+z-1)-3(1-z)(2x+3z-3)-v\Delta\right].
\eea
\end{subequations}
\end{widetext}
It is worth noting that the equations of the system \Ref{dynsys} are not
independent, because there exists the following dependence between the variables
$x$, $z$, and $v$:
\beq\label{constr_xzv}
z v^N(1-x-z)=-6^N(8\pi)^{\frac{2-N}{2}} V_0\kappa x^{\frac{N+2}{2}}(2x+2z-3)^{\frac{2-N}{2}}.
\eeq
Since the above relation is too complicated, in practice we solve the system \Ref{dynsys} straightforwardly, and then exclude surplus solutions.

%%%%%%%%%%%%%%%%%%%%%%%%%%%%%%%%%%%%%%%%%%%
\subsection{Stationary points, stability analysis,
and asymptotics}
%%%%%%%%%%%%%%%%%%%%%%%%%%%%%%%%%%%%%%%%%%%
In this section we study stationary points of the dynamical system \Ref{dynsys}
and perform a stability analysis of these points.
To find a stationary point $(x_0,z_0,v_0)$, we set $x_0'=z_0'=v_0'=0$ in Eqs.
\Ref{dynsys} and solve the resulting algebraic equations. Then, we investigate
its stability with respect to small perturbations $\delta x$, $\delta z$, and
$\delta v$ around $(x_0,z_0,v_0)$. Explicitly, we substitute
\beq
x=x_0+\delta x,\quad z=z_0+\delta z,\quad
v=v_0+\delta v
\eeq
into Eqs. \Ref{dynsys} and keep terms up to the first order in $\delta x$,
$\delta z$, $\delta v$. This leads to a system of first-order
differential equations
\beq
\frac{d}{d(\ln a)}\left(
\begin{array}{c}
 \delta x\\ \delta z\\ \delta v
\end{array}
\right)
={\cal M}
\left(
\begin{array}{c}
 \delta x\\ \delta z\\ \delta v
\end{array}
\right),
\eeq
where $\cal M$ is a $3\times 3$ matrix which depends on $(x_0,z_0,v_0)$. The
stability of the stationary point $(x_0,z_0,v_0)$ is determined by corresponding
eigenvalues $(\lambda_1,\lambda_2,\lambda_3)$ of $\cal M$. In particular,
if real parts of all eigenvalues are negative the point is stable (local sink),
if all real parts are positive the point is unstable being stable while integrating
in the opposite time direction (local source), if there are eigenvalues with different
signs of their real parts the point is a saddle.

In the table \ref{tab01} we enumerate all stationary points of the
dynamical system \Ref{dynsys}, briefly characterize their stability, and give
asymptotics for $a(t)$ and $\phi(t)$. It is necessary to stress that we only consider those points which satisfy the additional constraint \Ref{constr_xzv}. Below, let us discuss the stationary points in more detail.

\begin{table*}
\caption{\label{tab01} Stationary points of the dynamical system \Ref{dynsys}.}
%\begin{ruledtabular}
\begin{tabular}{cclclcl}
\hline\hline
\textbf{No} &\quad & \textbf{Stationary point} &\quad & \textbf{Stability} &\quad & \textbf{Conditions of an existence}\\
\hline
1. &\quad & $x=0$, $y=1$, $z=0$, $v=0$ &\quad & Unstable node &\quad & $\forall N$, $\kappa<0$, $t\rightarrow t_0$\\
2. &\quad & $x=\frac12$, $y=-\frac12$, $z=1$, $v=0$ &\quad & Complex type &\quad & $0<N<2$, $\kappa>0$, $t\rightarrow\infty$\\
3. &\quad & $x=1$, $y=0$, $z=0$, $v=0$ &\quad &  Saddle point &\quad &
$V(\phi)\equiv 0$, $\forall\kappa$, $t\rightarrow \infty$\\
4. &\quad & $x=0$, $y=-\frac12$, $z=\frac32$, $\textstyle
v=\frac{12}{3N+2}$ &\quad & Stable node &\quad & $N>2$, $\forall\kappa$, $t\rightarrow t_0$\\
5. &\quad & $x=\frac32$, $y=-\frac12$, $z=0$, $v=-3$ &\quad & Unstable node &\quad & $0<N<2$, $\kappa>0$, $t\rightarrow-\infty$\\
\hline\hline
\end{tabular}
%\end{ruledtabular}
\end{table*}

\subsubsection{The stationary point $x=0$, $y=1$, $z=0$, $v=0$.}
In this case the eigenvalues read
$$\textstyle
\lambda_1=3,\ \lambda_2=3,\ \lambda_3=\frac32.
$$
Since all eigenvalues are positive, this
point represents an unstable node for any $\kappa$, $V_0$, and $N$.
Substituting $x=0$, $y=1$, $z=0$, and $v=0$ into Eq. \Ref{Y}, we find that
$Y=-\frac32$ at the stationary point. Then, using the definition $Y=\dot H/H^2$,
we can obtain an asymptotical form of $a(t)$:
\beq\label{as-a-1}
a(t)=a_0(t-t_0)^{2/3}.
\eeq
An asymptotic for $\phi(t)$ can be found from the relation
%$y=-8\pi\kappa\dot\phi^2/2(1+8\pi\kappa\dot\phi^2)$
$y=-\frac{8\pi\kappa\dot\phi^2}{2(1+8\pi\kappa\dot\phi^2)}$ (see Eq. \Ref{def:xyzv}); putting $y=1$ into the latter yields
\beq\label{as-p-1}
\phi(t)=\phi_0+\phi_1(t-t_0),
\eeq
where $\phi_1^2=-\frac{1}{12\pi\kappa}$ and $\kappa<0$.
%Note that this asymptotic can exist only if $\kappa<0$.
Additionally one can substitute the asymptotics \Ref{as-a-1} and \Ref{as-p-1}
into Eqs. \Ref{def:xyzv} and check that $x\to0$, $y\to1$, $z\to0$, and
$v\to0$ as $t\to t_0$, where $t_0$ is an initial moment of time.
Note the same asymptotic was also obtained in the model with
$V(\phi)\equiv 0$ \cite{Sus:2009}.

\subsubsection{The stationary point $x=\frac12$, $y=-\frac12$, $z=1$, $v=0$.}
%Only one point $x=-y=\frac12$, $z=1$, $v=0$ for $N>2$ of this line satisfies to the constraint $v^N=-\frac{V_0\kappa 6^N x^{\frac{N+2}{2}}{(-8\pi(3-2x-2z))}^{\frac{2-N}{2}} }{(1-x-z)z}$.
In this case the eigenvalues are
$$\textstyle
\lambda_1=0,\ \lambda_2=0,\ \lambda_3=-3.
$$
Since two of these eigenvalues are equal to zero, one needs an additional study
to characterize a stability of the stationary point. In the next section we will
discuss this problem using a numerical analysis.

To find an asymptotic for $a(t)$, we take into account that $\frac{y}{x}=-1$ at
the stationary point. By using the definitions \Ref{def:xyzv} for $x$
and $y$, we can obtain $H^2=\frac{1}{3\kappa}$, which is possible only if
$\kappa>0$. Now, the asymptotic for $a(t)$ reads
\beq\label{as-a-2}
a(t)=a_0 e^{\frac{t}{\sqrt{3\kappa}}}.
\eeq

Analogously, to find an asymptotic for $\phi(t)$, we use the relation
$\frac{z}{x}=2$. Substituting Eqs. \Ref{def:xyzv} into this relation and
integrating, we can obtain
\begin{equation}\label{as-p-2}
\phi(t)=\phi_0 t^{\frac{2}{2-N}},
\end{equation}
where $\phi_0=\left[\frac12 (2-N)\sqrt{V_0}\right]^{\frac{2}{2-N}}$.
Additionally, substituting the asymptotics \Ref{as-a-2} and \Ref{as-p-2} into
\Ref{def:xyzv}, one can check that $x\to \frac12$, $y\to -\frac12$,
$z\to1$, and
$v\to0$ in the limit $t\to\infty$ only if $0<N<2$.

\subsubsection{The stationary point $x=1$, $y=0$, $z=0$, $v=0$.}
In this case the eigenvalues are
$$
{\lambda}_1=-6,\ {\lambda}_2=6, \ {\lambda}_3=0.
$$
Since two of three eigenvalues have opposite signs, this stationary point is a
saddle point for any $\kappa$, $V_0$, and $N$. From Eq. \Ref{Y} we find $Y=-3$.
Then, from the definition \Ref{def:betaXY} we obtain an asymptotic for $a(t)$ as
follows
\begin{equation}\label{as-a-4}
a(t)=a_0 t^{1/3}.
\end{equation}
From the definitions \Ref{def:xyzv} we conclude that $\dot\phi^2\to0$ if $y\to
0$, and $8\pi\dot\phi^2/6H^2\to 1$ if $x\to 1$. Integrating the relation
$8\pi\dot\phi^2/6H^2=1$ and using Eq. \Ref{as-a-4}, we can obtain an asymptotic
for $\phi(t)$:
\beq\label{as-p-4}
\textstyle
\phi(t)=\phi_0+\phi_1\ln t.
\eeq
with $\phi_1^2=\frac{1}{12\pi}$. Additionally, one should check that the
relations \Ref{def:xyzv} provide necessary limiting values. Substituting the
asymptotics \Ref{as-a-4} and \Ref{as-p-4} into \Ref{def:xyzv}, we can see that
$x\to 1$, $y\to 0$, and $v\to 0$ at $t\to\infty$. However, it is worth noting
that the necessary limit $z=0$ is only fulfilled if $V_0=0$, i.e.
$V(\phi)\equiv 0$. In this case we obtain the well-known solution for  a
minimally coupled (i.e. $y=0$ or, equivalently, $\kappa=0$) massless (i.e.
$V=0$) scalar field \cite{Sus:2009}.

%This point represents the well-known regime for a minimmaly
%coupled massless scalar field (since $y=0$ the regime corresponding to this
%point exists in the theory of minimally coupled scalar field).

\subsubsection{The stationary point $x=0$, $y=-\frac12$, $z=\frac32$, $v=\frac{12}{3N+2}$.}
In this case the eigenvalues are
$$
\lambda_1=-\frac{6(N-2)}{3N+2},\ \lambda_2=-\frac{6(N+2)}{3N+2}, \ \lambda_3=-6.
$$
Note that $\lambda_2$ and $\lambda_3$ are negative, while a sign of $\lambda_1$
depends on $N$. Namely, (i) $\lambda_1<0$ if $N>2$, and hence the stationary
point is an attractive node; (ii) $\lambda_1>0$ if $N<2$, and the stationary
point is a saddle point; (iii) $\lambda_1=0$ if $N=2$, and one needs an
additional study to characterize a stability of the stationary point.
Assume that $N\not=2$. Now, using Eq. \Ref{Y}, we find $Y=\frac{3(N-2)}{2N+2}$. The corresponding
asymptotics for $a(t)$ are as follows:
\beq\label{as-a-6}
a(t)=a_0(t-t_0)^{-\frac{3 N+2}{3(N-2)}}.
\eeq
In order to obtain the asymptotical behavior of $\phi(t)$, we use the
definition $v=\frac{\dot\phi}{\phi H}$. In our case we find $\frac{\dot\phi}{\phi
H}=\frac{12}{3N+2}$. Then, integrating gives
\beq
\phi=C a^{\frac{12}{3N+2}},
\eeq
where $C$ is a constant of integration.
Substituting Eq. \Ref{as-a-6} into the latter relation yields
\beq\label{as-p-6}
\phi(t)=\phi_0(t-t_0)^{-\frac{4}{N-2}}
\eeq
Additionally, substituting the asymptotics \Ref{as-a-6} and \Ref{as-p-6} into
\Ref{def:xyzv}, one can check that $x\to 0$, $y\to -\frac12$, $z\to\frac32$, and
$v\to\frac{12}{3N+2}$ in the limit $t\to t_0$ only if $N>2$. It is clear that for $N>2$ this point
represents a Big Rip asymptotic.

%%%%%%%%%%%%%%%%%%%%%%%%%%
\subsubsection{The stationary point $x=\frac{3}{2}$, $y=-\frac{1}{2}$, $z=0$,
$v=-3$.}
In this case the eigenvalues are
$$
\lambda_1=3(2-N), \ \lambda_2=6, \ \lambda_3=3.
$$
Since two of three eigenvalues are positive, this point is unstable for any $\kappa$ and $V_0$. Namely, it is a saddle if $N>2$, or an unstable node if $N<2$. Using Eq. \Ref{Y}, we calculate $Y=0$, and hence the relation $Y=\frac{\dot H}{H^2}$ yields $H=const$. Now, using the other relation $\frac{y}{x}=-3\kappa H^2$, we obtain $H^2=\frac{1}{9\kappa}$, which is possible only if $\kappa>0$. The resulting asymptotic for $a(t)$ is
\begin{equation}\label{as5-a}
a(t)=a_0 e^{\frac{t}{\sqrt{9\kappa}}}.
\end{equation}
Substituting $v=-3$ and $H=\frac{1}{\sqrt{9\kappa}}$ into the relation $v=\frac{\dot{\phi}}{\phi H}$, we can obtain the asymptotic for $\phi(t)$:
\begin{equation}\label{as5-phi}
\phi(t)=\phi_0 e^{-\frac{t}{\sqrt{\kappa}}}.
\end{equation}
Taking into account the definition \Ref{def:xyzv}, we can see that
$\dot\phi^2\to\infty$ if $y\to-\frac12$. Hence the asymptotics \Ref{as5-a} and
\Ref{as5-phi} are realized at $t\to-\infty$.
Additionally, let us consider an asymptotical behavior of $z$. For the
power-law potential the definition \Ref{def:xyzv} gives
$
z=\frac{8\pi V_0\phi^N}{3H^2(1+8\pi\kappa\dot\phi^2)}.
$
Substituting the asymptotics \Ref{as5-a} and \Ref{as5-phi} into the latter
relation, we can see that $z\rightarrow 0$ at $t\to-\infty$ only if $0<N<2$.
Note that  the same asymptotic have also been  obtained in the model with
$V(\phi)\equiv 0$ \cite{Sus:2009}.

%%%%%%%%%%%%%%%%%%%%%%%%%%%%%%%%%%%%%%%%%%%
\section{Examples of cosmological scenarios}
%%%%%%%%%%%%%%%%%%%%%%%%%%%%%%%%%%%%%%%%%%%
In this section we examine some specific cosmological scenarios corresponding
to specific potential choices. Since we are mostly interesting in the inflation driven by nonminimal kinetic coupling, hereafter we will assume $\kappa>0$.

First, let us separate the equation for $\phi$ and
$H$. For this aim we resolve Eqs. \Ref{11cmpt} and \Ref{eqmocosm} with
respect to $\dot H$ and $\ddot\phi$ and then, using the constraints
\Ref{constrphigen} and \Ref{constralphagen}, we can eliminate
$\dot\phi$ and $H$ from respective equations and find
\begin{widetext}
\beq\label{phi2gen}
    \ddot\phi=\frac{-2\sqrt{3\pi}\dot\phi
    [1+8\pi\kappa\dot\phi^2-8\pi\kappa V(\phi)]
    \sqrt{[\dot{\phi}^2+2V(\phi)](12\pi\kappa\dot\phi^2+1)}
    -(12\pi\kappa\dot\phi^2+1)(4\pi\kappa\dot\phi^2+1)V_\phi}
    {1+12\pi\kappa\dot\phi^2+96\pi^2\kappa^2\dot\phi^4
    +8\pi\kappa V(\phi)(12\pi\kappa\dot\phi^2-1)},
\eeq
\beq\label{a2gen}
\dot H=\frac{-(1-3\kappa H^2)(1-9\kappa H^2)[
3H^2-8\pi V(\phi)]+
4\sqrt{\pi}\kappa H\sqrt{(1-9\kappa H^2)[3H
^2-8\pi V(\phi)]}\,V_\phi}
    {1-9\kappa H^2+54\kappa^2H^4-8\pi\kappa
V(\phi)(1+9\kappa H^2)}.
\eeq
\end{widetext}
We mention however that although the $\phi$-equation does not
contains $H$-terms, the $H$-equation in general contains
$\phi$-terms arising from the potential $V(\phi)$. For this reason, in practice we will construct a solution $H(t)$ by substituting $\phi$, found as a numerical solution of Eq. \Ref{phi2gen}, into \Ref{constralphagen}.

%%%%%%%%%%%%%%%%%%%%%%%%%%%%%%%%%%%%%%%%%%%
\subsection{Oscillatory asymptotic}
%%%%%%%%%%%%%%%%%%%%%%%%%%%%%%%%%%%%%%%%%%%
Asymptotical properties of Eqs. \Ref{phi2gen}, \Ref{a2gen} depend on asymptotical values of the derivative $\dot\phi$ and the scalar potential $V(\phi)$. First, let us suppose that corresponding
asymptotical values are sufficiently small, so that
$$
8\pi\kappa\dot\phi^2\ll 1, \quad 8\pi\kappa V(\phi)\ll 1.
$$
By neglecting corresponding terms, Eq. \Ref{phi2gen} takes the following approximate form:
\beq\label{aseqVgen}
    \ddot\phi=-2\sqrt{3\pi}\dot\phi \sqrt{\dot{\phi}^2+2V(\phi)}-V_\phi.
\eeq
It is worth noting that this equation does not contain $\kappa$ and has the same form as in the theory of the usual minimally coupled scalar field. It has well-know asymptotics which are  represented as damped oscillations. In the particular case of the quadratic potential $V(\phi)=V_0\phi^2$ one has \cite{Star}
%\beq\label{aseqV2}
%    \ddot\phi=-2\sqrt{3\pi}\dot\phi \sqrt{\dot{\phi}^2+2V_0\phi^2}-2V_0\phi.
%\eeq
%Substituting $\phi(t)=At^{-1}\sin\omega t$ into Eq. \Ref{aseqV2} and keeping terms of order $t^{-2}$, we obtain the following asymptotical solution:
\beq\label{damposcilphi}
\phi_{t\to\infty}\approx \frac{\sin m t}{\sqrt{3\pi}\,m t},
\eeq
%which represents damped harmonic oscillations. Substituting this asymptotic into Eq. \Ref{constralphagen}, we can find the asymptotical behavior of the Hubble parameter $H(t)$:
and
\beq\label{damposcilH}
H_{t\to\infty}\approx H_{MD}(t)\,\left[1-\frac{\sin 2mt}{2mt}\right],
\eeq
where $m=\sqrt{2V_0}$ is a scalar mass and $H_{MD}(t)=2/3t$ is the Hubble parameter in the matter-dominated Universe
filled with nonrelativistic matter with $p\ll\rho$.
%Thus, the asymptotic \Ref{damposcilH} represents damped harmonic oscillations around $H_{MD}(t)$.

%In the general case with $V(\phi)=V_0|\phi|^N$ the equation \Ref{aseqVgen} reads
%\beq\label{aseqVN}
%    \ddot\phi=-2\sqrt{3\pi}\dot\phi \sqrt{\dot{\phi}^2+2V_0|\phi|^N}
%    -N V_0|\phi|^{N-1}\frac{\phi}{|\phi|}.
%\eeq
%It could be shown numerically that for all $N$ solutions of Eq. \Ref{aseqVN} represent damped oscillations. As the example, in Fig. \ref{f-oscillations} we show an exact solution of Eq. \Ref{aseqVN} with $N=3/2$.
%\begin{figure}[h]
%\begin{center}
%\includegraphics[width=7cm]{asphi.jpg}%
%\end{center}%
%\caption{\label{f-oscillations} Damped oscillations.}
%\end{figure}
%%The corresponding asymptotic for $H(t)$ has the form of damped oscillations -- generally anharmonic -- around the Hubble parameter $H_0(t)$ corresponding to the perfect fluid with the equation of state $p=w\rho$, where the EoS parameter $w$ depends on $N$ \cite{aaa}.

%Concluding this consideration, we can say that

%%%%%%%%%%%%%%%%%%%%%%%%%%%%%%%%%%%%%%%%%%%
\subsection{Exponential asymptotic}
%%%%%%%%%%%%%%%%%%%%%%%%%%%%%%%%%%%%%%%%%%%
Now, let us assume that the scalar field has an exponential
asymptotic: % at $t\to\pm\infty$:
\beq\label{asformphi}
\phi(t) \approx \phi_0 e^{\lambda t},
\eeq
where $\lambda>0$ if $t\to\infty$, and $\lambda<0$ if $t\to-\infty$.
In this case, asymptotically, $\phi\sim\dot\phi\sim\ddot\phi$.
Since asymptotical properties of $V(\phi)=V_0\phi^N$ depend on $N$, we will
consider different cases separately.

\vskip6pt $\mathbf{N<2}$.~In this case, asymptotically,
$
V(\phi)=V_0\phi^N\ll\dot\phi^2,\ V_\phi=NV_0\phi^{N-1}\ll\dot\phi.
$
Substituting the asymptotic \Ref{asformphi} into \Ref{phi2gen} and using the
relevant asymptotical properties, we can find
\beq
\lambda=-\frac{1}{\sqrt{\kappa}}.
\eeq
Since $\lambda=-1/\sqrt{\kappa}<0$, the corresponding asymptotic \Ref{asformphi} is carried out at the distant past, i.e. at $t\to-\infty$. Moreover, the requirement that $\lambda$ should be real yields $\kappa>0$.
Now, from Eq. \Ref{asformphi} we find
\beq\label{asphi1}
\phi_{t\to-\infty}\sim e^{-{t}/{\sqrt{\kappa}}}.
\eeq
Then, using the constraint \Ref{constralphagen}, we can obtain the asymptotic for
$H$:
\beq\label{asH1}
H_{t\to-\infty}\sim {1/\sqrt{9\kappa}}.
\eeq
This inflationary asymptotic corresponds to the stationary
point $5$ which is a local source for
phase trajectories. It should be emphasized that the standard inflation regime
for a minimally coulped scalar field does not share this property. Moreover, the exponential regime is highly unprobable during the contraction phase of the Universe and requires some special initial conditions \cite{Star}.
On the contrary, in the theory with nonminimal kinetic coupling all trajectories
in our numerical experiments (for full numerical results see below)
have reached this regime in a far past.

\vskip6pt $\mathbf{N=2}$.~The point $5$ does not exist for quadratic potential, so we provide a special analysis for this physically important case.
We have now $V(\phi)=V_0\phi^2\sim\dot\phi^2,\
V_\phi=2V_0\phi\sim\dot\phi$.
Using the asymptotic \Ref{asformphi}, we can find from Eqs. \Ref{phi2gen}, \Ref{constralphagen} the following asymptotical solutions:
\beq\label{asphi2}
\phi_{t\to\pm\infty}\sim
\exp\left[-\frac{t}{\sqrt{\kappa}}\,\frac{1-\mu}{\sqrt{1+2\mu}}\right],
\eeq
\beq\label{asH2}
H_{t\to\pm\infty}\sim \sqrt{\frac{1+2\mu}{9\kappa}},
\eeq
where $\kappa>0$ and $\mu$ is an auxiliary parameter, which can be found as a solution of the following equation:
\beq\label{cubicmu}
\kappa V_0=\frac{\mu(1-\mu)^2}{(1+2\mu)}.
\eeq
The latter is a cubic equation with respect to $\mu$. Generally, it has three roots $-0.5<\mu_1<0.25$, $0.25<\mu_2<1$, and $\mu_3>1$ provided $\kappa V_0<3/32$. For $\kappa V_0=3/32$ two roots are coinciding, so that $\mu_1=\mu_2=0.25$. In case $\kappa V_0>3/32$ the only root $\mu_3>1$ remains.
%It is physically reasonable to assume
Supposing that $\kappa V_0\ll 1$, we can easily obtain the approximate solution of Eq. \Ref{cubicmu}:
\beq
\mu_1\approx \kappa V_0,\quad \mu_2\approx 1-\sqrt{3\kappa V_0},\quad \mu_3\approx 1+\sqrt{3\kappa V_0}.
\eeq
Correspondingly, Eqs. \Ref{asphi2} and \Ref{asH2} lead to the following three asymptotics:
\bea
{\rm A1.} &&
\phi_{t\to-\infty}\sim \exp\left[-\frac{t}{\sqrt{\kappa}}(1-2\kappa V_0)\right],
\nonumber\\
&& H_{t\to-\infty}\sim \frac{1}{\sqrt{9\kappa}}(1+\kappa V_0),\\
{\rm A2.} &&
\phi_{t\to\infty}\sim \exp\left[t\sqrt{3 V_0}\right],
\nonumber\\
&& H_{t\to\infty}\sim \frac{1}{\sqrt{3\kappa}}\left(1+\sqrt{\frac{\kappa
V_0}{3}}\right),\\
{\rm A3.} &&
\phi_{t\to-\infty}\sim \exp\left[-t\sqrt{3 V_0}\right],
\nonumber\\
&& H_{t\to-\infty}\sim \frac{1}{\sqrt{3\kappa}}\left(1-\sqrt{\frac{\kappa
V_0}{3}}\right).
\eea
Note that the asymptotics A1 and A3 are realized at the distant past
$t\to-\infty$, while the asymptotic A2 is carried out at the future
$t\to\infty$.

\vskip6pt $\mathbf{N>2}$.~In this case $V(\phi)=V_0\phi^N\gg\dot\phi^2,\
V_\phi=NV_0\phi^{N-1}\gg\dot\phi$, and one can check straightforwardly that
$\phi(t)\sim e^{\lambda t}$ cannot be an asymptotic of Eq. \Ref{phi2gen} at
$t\to\pm\infty$.

%%%%%%%%%%%%%%%%%%%%%%%%%%%%%%%%%%%%%%%%%%%
\subsection{Cosmological model with $V(\phi)=V_0|\phi|^{3/2}$}
\begin{figure}[ht]
\begin{center}
\parbox{7.5cm}{\includegraphics[width=7.5cm]{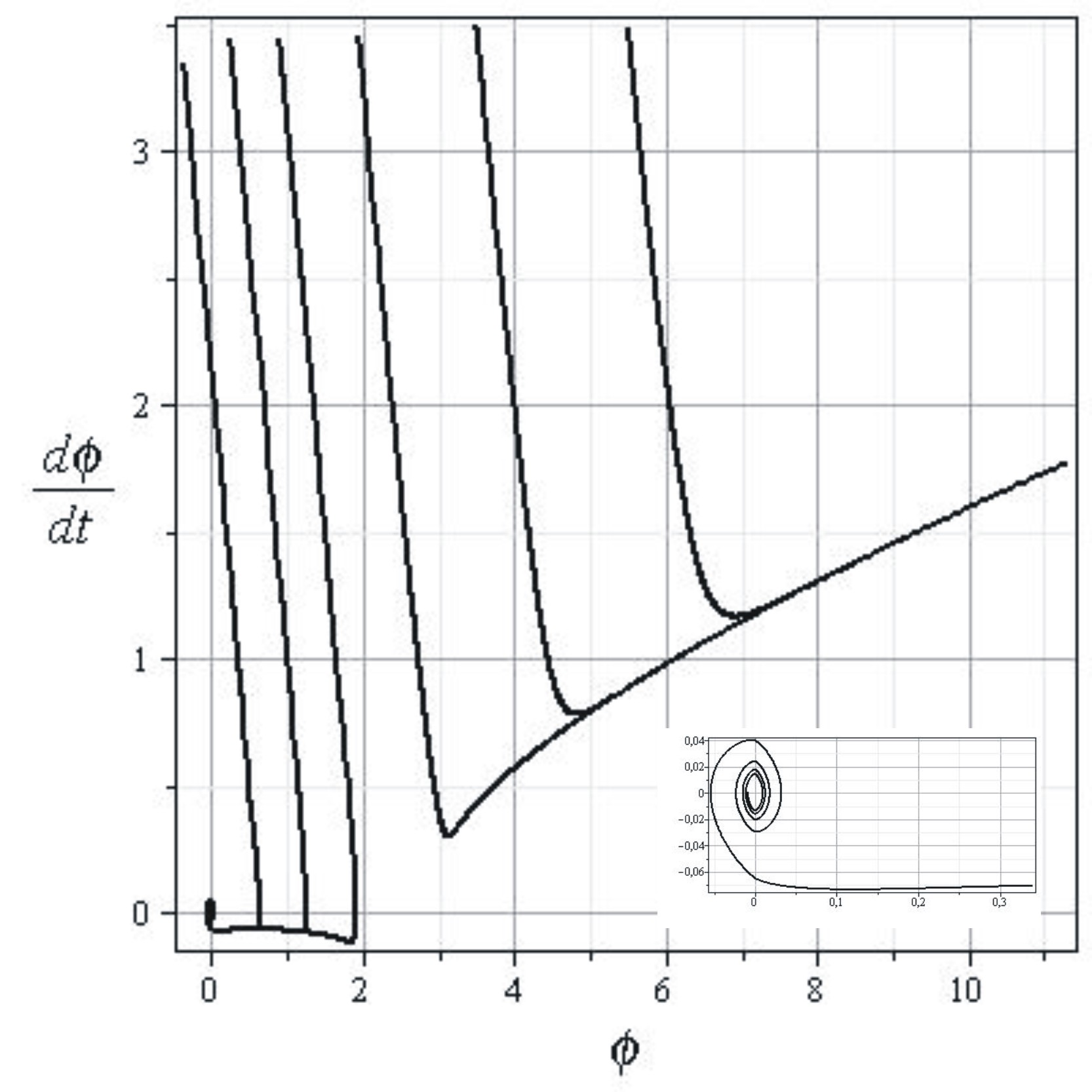}\\~~~~~~~~(a)}\\%
\parbox{7.5cm}{\includegraphics[width=7.5cm]{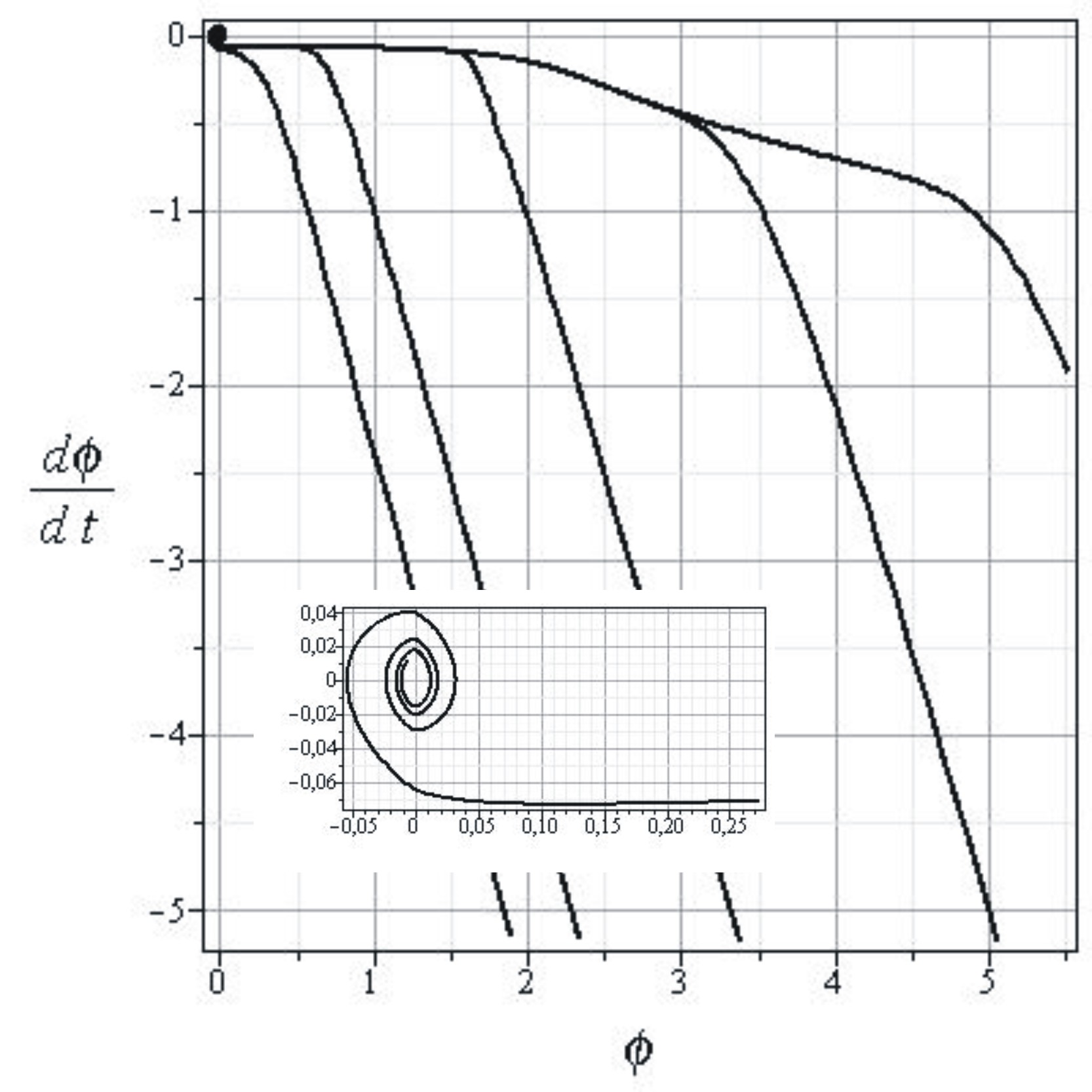}\\~~~~~~~~(b)}%
\end{center}%
\caption{\label{figP-N32} Phase diagrams for the scalar field $\phi(t)$ are presented for the coupling parameter $\kappa=0.1$ and the potential $V(\phi)=V_0|\phi|^{3/2}$ with $V_0=0.1$. The solutions are constructed for initial conditions $\phi(0)=\dot\phi(0)=\{0.5,1,1.5,2.3,3.5,5\}$ [plot (a)], and $\phi(0)=-\dot\phi(0)=\{0.1, 1, 2.5, 5, 7.5\}$ [plot (b)]. Phase trajectories in the vicinity of zero are shown separately on small plots.}
\end{figure}

\begin{figure}[ht]
\begin{center}
\parbox{7.5cm}{\includegraphics[width=7.5cm]{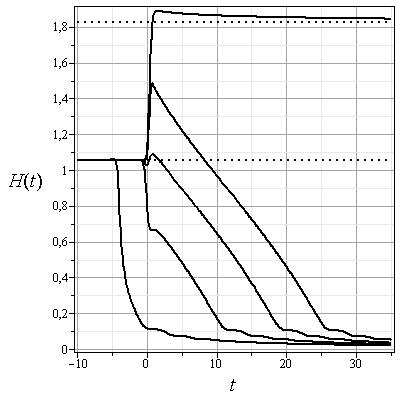}\\~~~~~~~~(a)}\\%
\parbox{7.5cm}{\includegraphics[width=7.5cm]{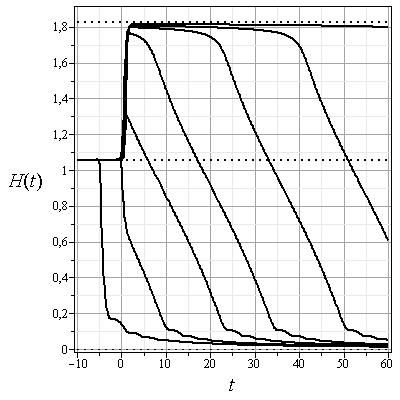}\\~~~~~~~~(b)}%
\end{center}%
\caption{\label{figH-N32} Graphs of $H(t)$ are presented for the coupling parameter $\kappa=0.1$ and the potential $V(\phi)=V_0|\phi|^{3/2}$ with $V_0=0.1$. The solutions are constructed for initial conditions $\phi(0)=\dot\phi(0)=\{0.04, 0.5, 1, 1.5, 3.5\}$ [plot (a)], and $\phi(0)=-\dot\phi(0)=\{0.05, 1, 2.5, 10, 100, 500, 5000\}$ [plot (b)]. The lower and upper dotted line show the asymptotics $1/\sqrt{9\kappa}$ and $1/\sqrt{3\kappa}$, respectively.}
\end{figure}

%%%%%%%%%%%%%%%%%%%%%%%%%%%%%%%%%%%%%%%%%%%
Let us consider the specific choice for the scalar potential. We start with the $N<2$ case. Namely, we assume that $N=\frac32$, so that
\beq
V(\phi)=V_0|\phi|^{3/2}.
\eeq
In order to present the cosmological scenario in this case more transparently,
we perform a numerical elaboration of the model given by Eqs. \Ref{phi2gen} and \Ref{constralphagen}. The numerical results are presented in Figs.
\ref{figP-N32} and \ref{figH-N32}.

%\begin{widetext}
%\end{widetext}

As was shown in the previous section, a typical trajectory starts with the inflationary regime \Ref{asphi1}-\Ref{asH1}. The numerical analysis gives that there exist two possibilities for the final fate of the trajectory: either it reaches the near-Einstein regime with small and dumping oscillation of the scalar field, or the evolution finishes in the second inflation described by Eqs. \Ref{as-a-2}-\Ref{as-p-2}. The latter leads to an eternal inflation, so it
suffers from the graceful exit problem. Our numerical data show that for both
$\kappa$ and $V_0$ being below unity the oscillatory regime dominates in the
future evolution, avoiding any difficulties with the graceful exit. In Fig. \ref{figP-N32}
a family of trajectories has been plotted for $\kappa=0.1$, $V_0=0.1$. Most
of trajectories ends with scalar field oscillations. What is
interesting, trajectories with big enough initial values of $\dot\phi$ plotted in Fig. \ref{figH-N32}b fall into oscillation not directly, but passing through a temporal second inflation phase. This property once more indicates the existence of rather complicated dynamics
in the vicinity of the point $2$.  This point acts first as an attractor, and then as
a repeller, giving the desired exit from inflation. Such the dynamical behavior is
similar for the standard inflation case \cite{Grishchuk} (for a recent development see, for example, \cite{Arefeva}, without, of course, any preceding inflation phase).

The qualitatively different cosmological behavior is represented by the other set of initial conditions plotted in Fig. \ref{figH-N32}a. Here we can see a trajectory which transits from the initial inflationary regime into the secondary one which never ends. We come to the conclusion that for some initial data (the trajectory 6 corresponds to big enough initial
values of the scalar field and its time derivative) the point $2$ can be stable. A complete description of the point $2$ requires a future work, here we only indicate that for
reasonably small values of $\kappa$ all trajectories with not so big initial
values of $\dot\phi$ do not enter an eternal secondary inflation.

We should also stress an important difference between standard inflation and
inflation described by the point $5$ of the present model. In the standard
inflation initial conditions are almost completely erased. In the inflation
under consideration the behavior of the scalar field is different from the usual
slow-roll regime, and the value of $\dot\phi$ at the end of inflation can be
different and depends on initial conditions. This leads to different fate of
trajectories after inflation, which can be clearly seen in the Fig. \ref{figH-N32} where
there are trajectories falling into oscillatory regime directly after first
inflation, trajectories reaching oscillation after transient second inflation
and trajectories never leaving the second inflation.

%%%%%%%%%%%%%%%%%%%%%%%%%%%%%%%%%%%%%%%%%%%
\subsection{Cosmological model with $V(\phi)=V_0\phi^{2}$}
\begin{figure}[ht]
\begin{center}
\parbox{7.5cm}{\includegraphics[width=7.5cm]{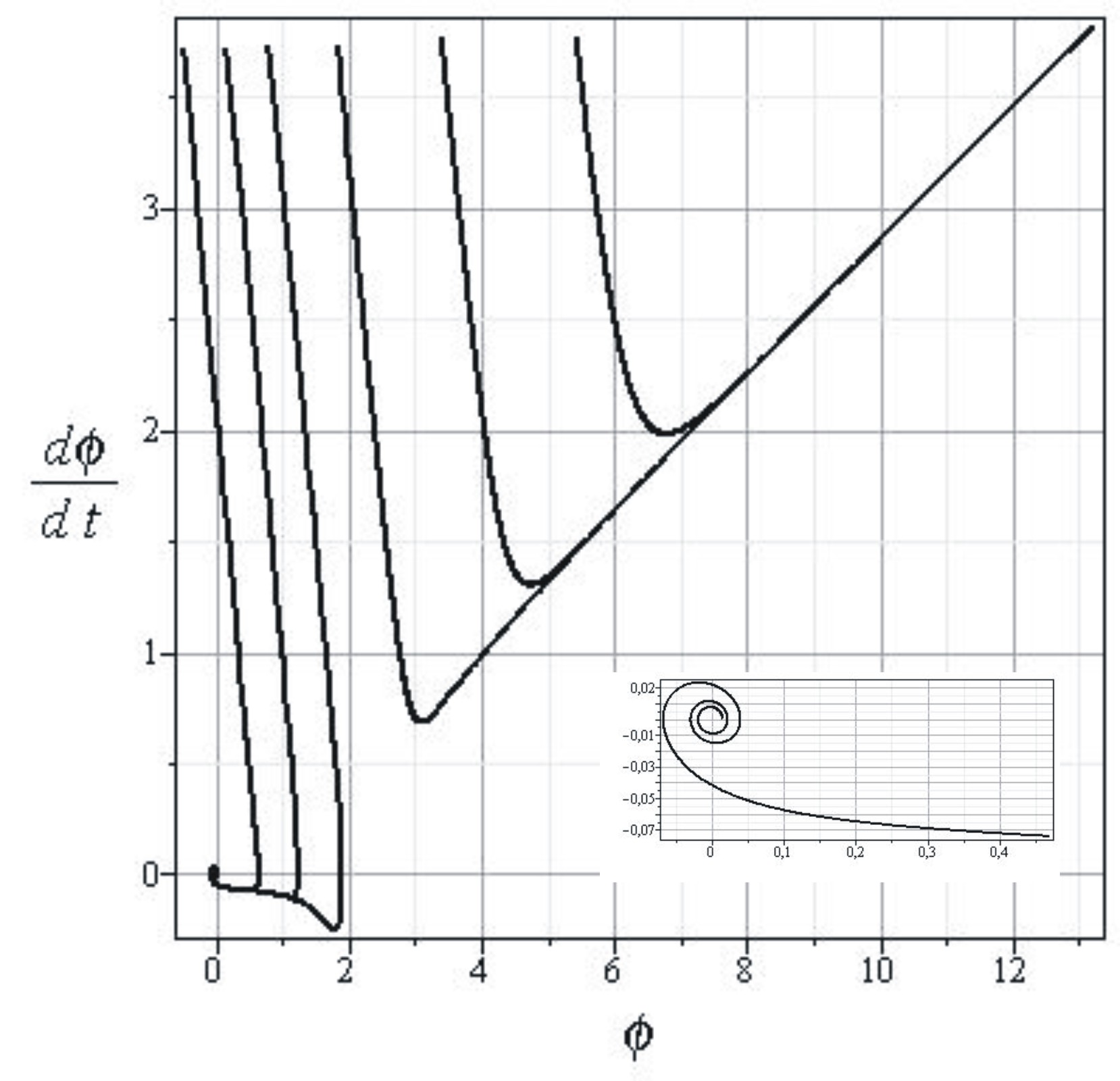}\\~~~~~~~(a)}\\%
\parbox{7.5cm}{\includegraphics[width=7.5cm]{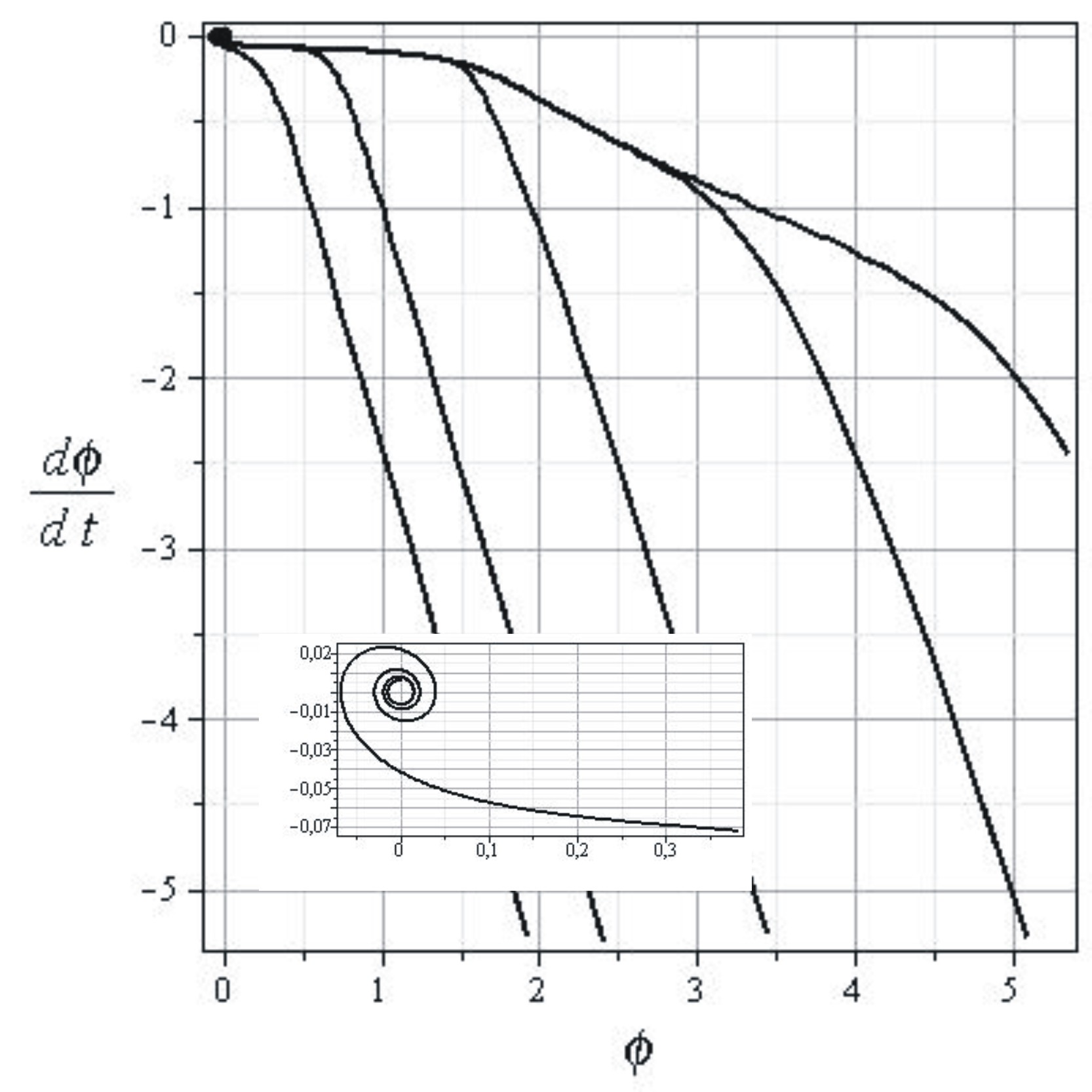}\\~~~~~~~(b)}
\end{center}%
\caption{\label{figP-N2} Phase diagrams for the scalar field $\phi(t)$ are presented for the coupling parameter $\kappa=0.1$ and the potential $V(\phi)=V_0\phi^{2}$ with $V_0=0.1$. The solutions are constructed for initial conditions $\phi(0)=\dot\phi(0)=\{0.5,1,1.5,2.3,3.5,5\}$ [plot (a)], and $\phi(0)=-\dot\phi(0)=\{0.1, 1, 2.5, 5, 7.5\}$ [plot (b)]. Phase trajectories in the vicinity of zero are shown separately on small plots.}
\end{figure}

\begin{figure}[ht]
\begin{center}
\parbox{7.5cm}{\includegraphics[width=7.5cm]{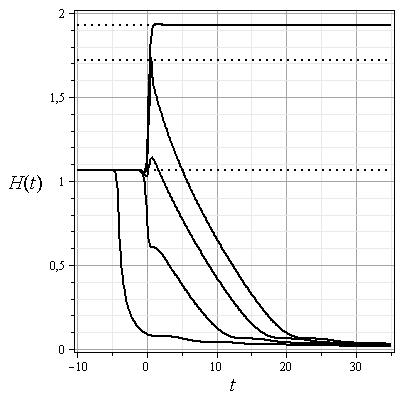}\\~~~~~~~(a)}\\%
\parbox{7.5cm}{\includegraphics[width=7.5cm]{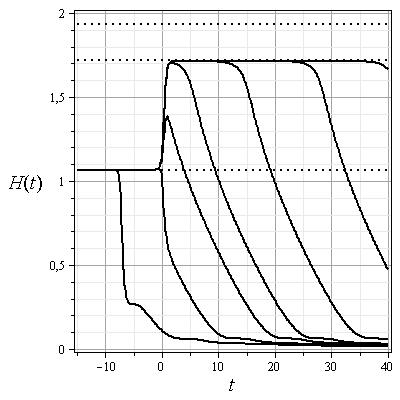}\\~~~~~~~(b)}%
\end{center}%
\caption{\label{figH-N2} Graphs of $H(t)$ are presented for the coupling parameter $\kappa=0.1$ and the potential $V(\phi)=V_0\phi^{2}$ with $V_0=0.1$. The solutions are constructed for initial conditions $\phi(0)=\dot\phi(0)=\{0.04, 0.5, 1, 1.5, 3.5\}$ [plot (a)], and $\phi(0)=-\dot\phi(0)=\{0.05, 1, 2.5, 10, 300, 30000, 3000000\}$ [plot (b)]. The dotted lines show the asymptotics $H_{t\to-\infty}\approx 1/\sqrt{9\kappa}(1+\kappa V_0)$ [lower line],  $H_{t\to\infty}\approx 1/\sqrt{3\kappa}\left(1-\sqrt{{\kappa V_0}/{3}}\right)$ [middle line], and $H_{t\to\infty}\approx 1/\sqrt{3\kappa}\left(1+\sqrt{{\kappa V_0}/{3}}\right)$ [upper line].}
\end{figure}

%%%%%%%%%%%%%%%%%%%%%%%%%%%%%%%%%%%%%%%%%%%
We remind a reader that in this case there are in general one or three exponential asymptotics depending on the value of the product $\kappa V_0$. In the latter case in all our numerical simulations the asymptotic (A1) is a local source. After this inflation ends the trajectory can go either to the asymptotic (A2) or to oscillatory regime. Our numerical results show that the asymptotic (A2) appears to be stable and the only possible way to reach ``graceful'' exit is to avoid it. On the other hand, the third asymptotic (A3) can provide a transient inflation phase
for appropriate initial conditions (see Figs. \ref{figP-N2} and \ref{figH-N2}).
Note that the case of single root is not favorable for inflationary scenario, because the single asymptotic appears to be stable and does not allow an exit from inflation.

Since for $N \ge 2$ the early time inflationary regime is absent, we do not consider
this case in the present paper. For such potentials the point $4$ is stable, so we can
suggest that the dynamics is dominated by a phantom-like behavior. We leave studies of non-inflationary regimes in the model under consideration to a future work.

%----------------------------------------------------------------
\section{Conclusions}
%----------------------------------------------------------------
We have considered cosmological dynamics for the FRW Universe filled with a
scalar field with kinetic coupling in the action \Ref{action}. One of the most
intriguing  feature of this model found earlier \cite{Sus:2009, SarSus:2010, Sus:2012} is existence of inflationary
behavior at early time in the case of zero or constant potential of the scalar field, i.e.
solely due to the coupling. This regime exists only for positive coupling
constant $\kappa$. In the present paper we study influence of nonzero scalar
field potential (for a negative $\kappa$ inflationary regime is absent for zero
potential, and nonzero potential leads to inflation qualitatively the same as
in the case of minimally coupled scalar field \cite{Tsujikawa}).

We have found that for the case of the quadratic potential, most interesting with the physical point of view,  the
inflationary regime exists for appropriate values of scalar field mass and
coupling constant. As for other power-law potentials, using theory of dynamical
system methods, we have found two other stable asymptotic regimes. One regime
leads to big rip singularity, and exists for potentials steeper than the
quadratic one. In this case the inflationary regime does not exist, so steep
potentials destroy the scenario of Ref.\cite{Sus:2012}.

On the other hand, for potentials which are more sloping than the quadratic one the
inflationary regime appears to be exactly the same as for zero/constant potential.
However, a new stable asymptotic regime appears which represents exponential
expansion and power-law increase of the scalar field. From the viewpoint of
expansion dynamics, it is an eternal inflation, so if the initial inflation ends by
reaching this regime, actual exit from inflation is absent. This is a danger for
this model. Our numerical study shows, however, that for wide range of
parameters of the theory a trajectory which exits from initial
inflation typically  does not reach the eternal secondary inflation regime, and scalar
field finally falls into oscillations.

In a summary, the scenario of initial inflation driven by the nonminimal kinetic coupling survives for a wide range of parameters provided the scalar potential is not steeper than the quadratic one.

%----------------------------------------------------------------
\section*{Acknowledgments}
%----------------------------------------------------------------
We are grateful to A.A.~Starobinsky for usual discussions.
The work was supported in part by the Russian Foundation for Basic Research grants Nos. 11-02-01162 and 11-02-00643.
%----------------------------------------------------------------

\end{document}